# Enhanced interfacial charge transfer by Z-scheme in defect-mediated ZnO-CdS nanocomposite with rGO as a solid-state electron mediator for efficient photocatalytic applications

Arun Murali [1], Rahulkumar Sunil Singh [1], Michael L Free [1] and Prashant K Sarswat [1*]

[1] Department of Materials Science and Engineering, University of Utah, Salt Lake City, Utah, 84112, U.S.A
* Correspondence: to whom correspondence should be addressed; E-mail: prashant.sarswat@utah.edu; Tel: +1-801-820-6919

**Abstract:** ZnO-based photocatalysts are widely investigated photocatalytic materials for pollutant degradation due to their low cost, abundance, and eco-friendly characteristics. However, the effectiveness of its photocatalytic properties is limited by inherent challenges such as a wide bandgap, photocorrosion, and rapid recombination of photogenerated charge carriers. In order to overcome these limitations observed in traditional ZnO photocatalysts and enhance their photocatalytic properties, an alternative approach has been proposed. This study introduces an oxygen defects-mediated Z-scheme mechanism for charge separation in the heterojunction by coupling $O_v$-ZnO with CdS, alongside the incorporation of rGO as an electron mediator. This mechanism aims to enhance the photostability and visible-light-induced photocatalysis properties of ZnO. Our work focuses on the development and characterization of trinary $O_v$-ZnO-rGO-CdS pho-to-catalysts, aiming to enhance their photocatalytic properties for efficient energy conversion and environmental applications. To characterize the trinary $O_v$-ZnO-rGO-CdS photocatalysts, we employed a range of characterization techniques, including X-ray diffraction, Raman spectroscopy, UV-Visible spectroscopy, Electrochemical impedance spectroscopy, Photoluminescence spectroscopy, and X-ray photoelectron spectroscopy. This approach not only provides insights into the defect-dependent interfacial mechanism in heterostructure nanocomposites but also opens promising possibilities for developing high-performance ZnO-based photocatalysts for energy conversion and environmental applications.

## 1. Introduction

In recent times, the escalating pace of industrialization has brought about significant global concerns regarding environmental degradation and energy sustainability [1]. A potential approach for addressing these issues is to develop effective renewable energy generation systems that can decrease our dependency on fossil fuels and consequently reduce greenhouse gas emissions [2]. Photocatalysis is one of the emerging green technologies that has gained significant attention since it can directly transform solar energy into chemical energy by mimicking natural photosynthesis, produce hydrogen from water, degrade organic pollutants, and reduce $CO_2$ into organic fuels [3, 4]. However, photocatalysts with a wide absorbance range, long-term stability, high charge-separation efficiency, and strong redox activity are required for the aforementioned applications. Generally, it is challenging for single-component photocatalysts to concurrently satisfy all of these requirements. Designing suitable heterogeneous photocatalytic systems can resolve this challenge [5]. The type-II heterojunction, and Z-scheme systems have been widely investigated in the last few decades to improve the photocatalytic efficiency [6, 7]. In a type-II heterojunction, the photoinduced electrons transfer to a less negative conduction band (CB) while the holes move to a less positive valence band (VB), leading to their lower redox ability [8]. Z-scheme system resembles to natural photosynthesis process in plants and involves two steps of photoexcitation [9]. In the Z-scheme, the photocatalyst possesses with similar staggered band structure configuration, but the charge transfer pathway is completely different [10]. Z-Scheme mode of the charge carrier pathway incorporates only the electrons and holes with strong redox ability to drive photocatalytic oxidation and reduction reactions, making it more desirable over type-II heterojunction [11]. In a traditional Z-scheme system photocatalyst, redox-mediator induced back reactions can occur, leading to a low photoconversion efficiency [12]. The selection of a proper electron conductor is crucial for efficient photocatalysts in an all-solid Z-scheme. Noble metals, such as Au, Ag, and Cu nanoparticles (NPs), have been used as excellent electron mediators. Recently, emerging carbonaceous materials, represented by reduced graphene oxide (rGO) and carbon dots (CDs) have been used as solid-state electron mediators [13]. The large surface area, good electric conductivity, and tunable band gap by altering the reduction rates of graphene oxide offer plenty of pathways and sites for Z-Scheme charge flow and recombination [14, 15]. Introducing oxygen vacancies is an effective strategy to preserve the intrinsic crystal structure of ZnO and extend the visible-light absorption [16]. Despite this, the serious photocorrosion in ZnO-based photocatalyst hinders the photocatalytic performance [17]. Among the various visible-light photocatalysts, CdS is a particularly favorable photocatalyst due to its band gap of 2.4 eV and high activity. CdS has a band gap of 2.4 eV [18]. A typical CdS material exhibits an absorption maxima at wavelength of ~ 514 nm.

To address these challenges, this idea was realized in the form of the trinary $O_v$-ZnO-rGO-CdS photocatalyst. In this configuration, rGO is strategically located at the interface of $O_v$-ZnO and CdS, played the role of mediating the recombination of electrons from the ZnO and holes from CdS. By elucidating the interfacial mechanisms and charge transfer pathways within the heterostructure nanocomposite, this study provides insights into the defect-dependent behavior and highlights the potential of Z-scheme configurations for improved photocatalytic performance. The approach also offers the potential to overcome the issue of photocorrosion commonly observed in ZnO-based photocatalysts, leading to improved stability and performance.

**2. Materials & methods**

*2.1. Materials*

Graphite powder with a particle size smaller than 20 μm, potassium permanganate ($KMnO_4$), sodium nitrate ($NaNO_3$), sodium thiosulphate ($Na_2S_2O_3·5H_2O$), zinc nitrate ($Zn(NO_3)_2·6H_2O$), cerium nitrate ($Ce(NO_3)_3·6H_2O$), hydrogen peroxide ($H_2O_2$), hydrochloric acid (HCl), and sulphuric acid ($H_2SO_4$) were acquired from Alfa Aesar. Ultrapure distilled water was utilized as a solvent and to prepare various solutions for the experiments. All the reagents employed in this study were of analytical grade.

*2.2. Experimental procedure*

Graphene oxide (GO) was synthesized using the Modified Hummers' method, following the previously described procedure [19]. In the synthesis of the ZnO-rGO nanocomposite, GO particles were dispersed in water and subjected to sonication, resulting in the formation of a yellow-brown solution. Subsequently, ZnO nanoparticles were gradually added to the solution under continuous stirring. The homogeneous suspension was then transferred to a Teflon-lined autoclave and subjected to hydrothermal treatment at 150 °C for 6 h [20]. The reduction of GO was indicated by the formation of a black-colored solution from the initial dark brown color, and the resulting material, ZnO-rGO, was thoroughly rinsed and dried, yielding a black solid. The hydrothermal treatment ensured the formation of a well-dispersed and stable composite material. The details of this hydrothermal process to synthesize ZnO-rGO nanocomposite are discussed in detail in previous literature [21]. Afterward, the ZnO-rGO sample was dipped in $NaBH_4$ solution to increase the density of oxygen vacancies. $O_v$-ZnO-rGO sample was then dispersed in DI water, and a certain volume of Cd(CH$_3$COO)$_2$ was added. It was then ultrasonicated, and Na$_2$S aqueous solution was added dropwise to the suspension [22]. The suspension was stirred, and the precipitate was washed, dried, and labeled as O$_v$-ZnO-rGO-CdS. The details of the working electrode preparation and the electrochemical measurements have been discussed in the previous work [21].

*2.3. Characterization*

The structural investigation was conducted using an X-ray diffractometer equipped with a Cu K$\alpha$1 radiation source ($\lambda$ = 1.5406 Å). X-ray photoelectron spectroscopy (XPS) was employed to determine the concentration of the Ce$^{3+}$ fraction, providing crucial information on the electronic structure and oxidation states of the materials. Raman spectra were recorded using a Raman spectrometer, employing an He-Ne laser, which provided information on the crystallinity, presence of defects, and structural changes within the samples. Photoluminescence spectra, recorded at room temperature using an excitation wavelength of 450 nm, provided details on the emission properties, energy band structure, and charge carrier dynamics of the materials. UV-Vis absorption spectra were obtained using a UV-3600 Shimadzu spectrophotometer, allowing for the investigation of the absorbance behavior of the materials across a wide range of wavelengths.

## 3. Results & Discussion

X-ray diffraction confirmed that the synthesized nanocomposite has patterns resembling to that of pure ZnO with a wurtzite phase [23]. The observed peaks at 2$\theta$ values of 32.3°, 34.8°, 36.3°, 47.6°, 56.6°, 62.9°, 66.4°, 67.9°, 69.1, 72.6, and 76.9° correspond to the (100), (002), (101), (102), (110), (103), (200), (112), (201), (004), and (202) crystal planes, respectively. These findings for ZnO are consistent with the standard card values for wurtzite-structured ZnO (JCPDS No. 36-1451) [24]. The comparison of the XRD patterns between the trinary O$_v$-ZnO-rGO-CdS nanocomposite and ZnO (Fig. 1a) demonstrated a positive shift in 2$\theta$ positions of the peaks, indicating a relatively smaller reduction in crystallite size compared to ZnO [25]. The XRD diffraction pattern of CdS revealed three distinct peaks, namely (111) at 26.5°, (220) at 43.9°, and (311) at 52.7°. These peak positions perfectly matched the standard data card for CdS (JCPDS card no. 89-0019) [26, 27]. However, distinct peaks corresponding to CdS and rGO were not detected in the diffraction pattern of O$_v$-ZnO-rGO-CdS, possibly due to the complete coverage of rGO surfaces by ZnO particles or relative thickness of these layers. This coverage hinders the observation of the expected reflection peaks of rGO at 2$\theta$ values of 24.6° and 43.3° [25]. Additional characterization technique, such as Raman spectroscopy was employed to further confirm the presence of rGO and CdS. In the Raman spectra, the characteristic D and G bands were identified and analyzed to gain insights into the structural properties and Raman shifting in both graphene oxide (GO) and the O$_v$-ZnO-rGO-CdS nanocomposite. Raman spectra showed that the intensity ratio of D (1330 cm$^{-1}$) and G (1540 cm$^{-1}$) bands corresponding to the sp$^3$ defects in carbon and sp$^2$ in carbon, respectively (Fig. 1b). This suggested the transformation of GO to rGO, which presented of much more unrepaired defects after the removal of a large amount of oxygen-containing functional groups [28]. The Raman D mode is activated when point-like defects are added to the planar sp$^2$ carbon lattice. The areal defect density can be assessed directly using its relative intensity in relation to the G mode. The absorbance spectra of ZnO, ZnO-rGO, and the O$_v$-ZnO-rGO-CdS composite (Fig. 1c) showed that ZnO and ZnO-rGO predominantly absorbed light with wavelengths below 400 nm due to their wide band gaps. The O$_v$-ZnO-rGO-CdS nanocomposite exhibited a noticeable shift, indicating band gap narrowing compared to C-ZnO. This was due to the presence of oxygen vacancies in O$_v$-ZnO-rGO-CdS, which created a defects-isolated level below the conduction band of ZnO. Consequently, the O$_v$-ZnO-rGO-CdS composite exhibited significantly enhanced visible-light absorption capacity. Figure 1(d) depicts the Nyquist plots, with the inset highlighting the shift in the Bode phase plots. The Nyquist plots (Fig. 1d) show that O$_v$-ZnO-rGO-CdS has lower electron-transfer resistance (diameter of the semicircle region) than ZnO-rGO and ZnO. This suggests that the heterojunction structure enhanced the electroconductivity of the nanocomposite. The XPS O 1s core level spectra (Fig. 1f) showed that it was quite asymmetric and can be described by the superimposition of three peaks by Gaussian distribution [29]. It is clear that a significant increase in the relative intensity of O$_v$ to O$_L$ component was observed in O$_v$-ZnO. The absorption edge exhibited a red shift for $O_v$-ZnO-rGO-CdS, implying band-gap narrowing due to the formation of oxygen vacancy related shallow donor levels [30]. It was observed from the PL spectroscopy (Fig. 1e) that the $O_v$-ZnO-rGO-CdS sample exhibited decreased emission intensity signifying an effective charge transport. $O_v$-ZnO-rGO-CdS sample displayed the highest electron-life time of 6.2 ms, followed by ZnO-rGO (4.4 ms), determined from the characteristic frequency ($\tau = \frac{1}{2\pi f_o}$). The extension of the emission lifetime in the case of inorganic NPs is occasionally thought to be caused by the lower frequency oscillation of the inorganic matrix, which is suggested by the correlation of emission lifetimes with the vibrational frequencies of the immediate environment. The observed elongation of emission lifetime in present case, suggested that the photoinduced electrons and holes, are more effectively separated in the heterojunction [31].

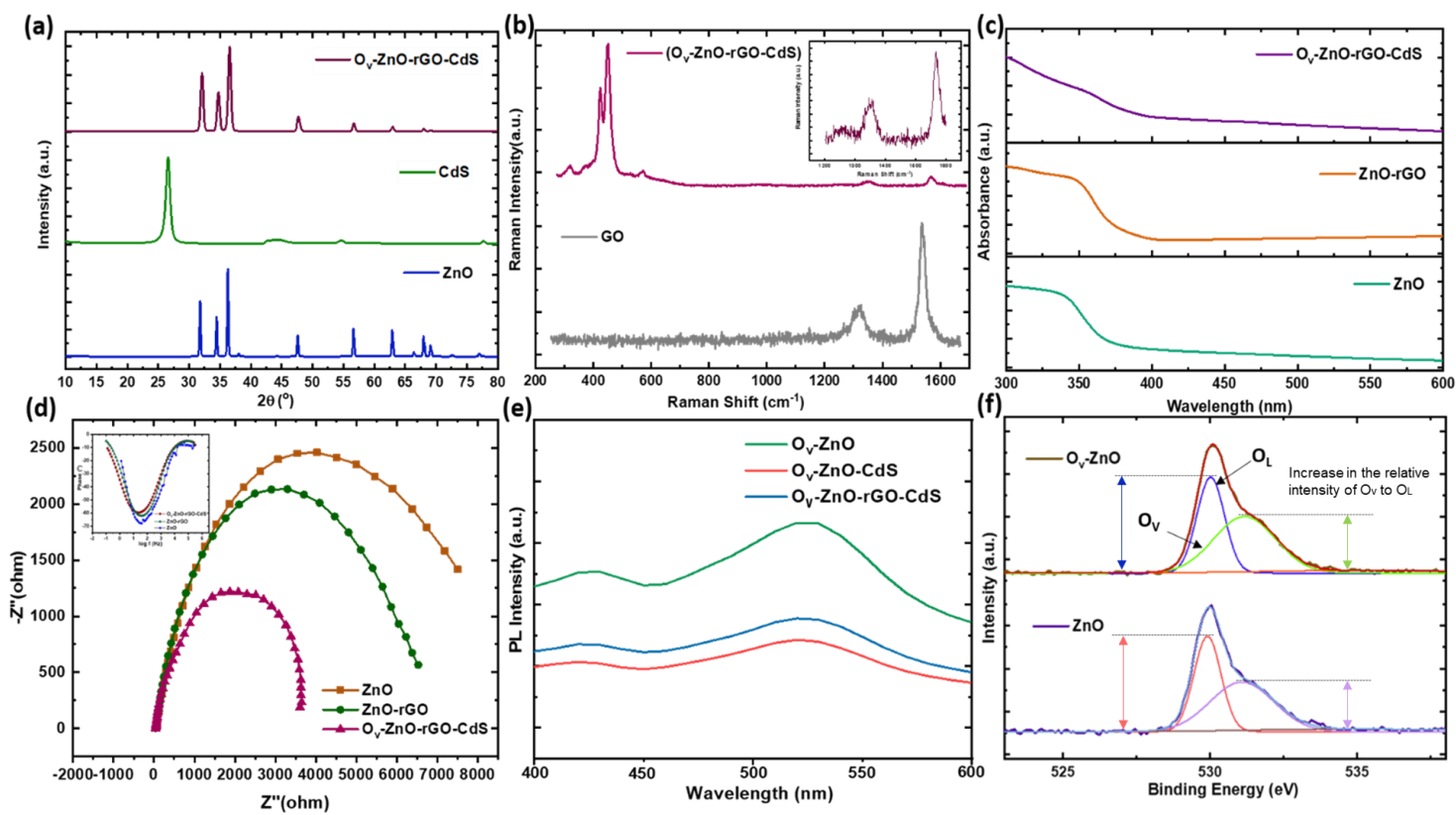

**Figure 1. (a)** XRD patterns of CdS, ZnO, and $O_v$-ZnO-rGO-CdS composites. **(b)** Raman spectra of GO and $O_v$-ZnO-rGO-CdS; inset shows the magnified D and G bands in $O_v$-ZnO-rGO-CdS. **(c)** Absorbance Absorbance spectra of the ZnO, ZnO-rGO, and $O_v$-ZnO-rGO-CdS nanocomposites. nanocomposites. **(d)** Nyquist plots and the inset shows the shift in the Bode phase plots. **(e)** PL spectra of the samples **(f)** The XPS O1s deconvoluted into two components with different binding energies.

The photocatalytic degradation of the nanocomposites was evaluated under visible-light irradiation using Rhodamine (RhB) dye. 55 % RhB was degraded after 100 min of irradiation for ZnO-rGO-CdS photocatalyst. When CdS was coupled with oxygen mediated ZnO, the degradation activity was significantly enhanced with a degradation of 94 %. The $O_v$-ZnO-rGO-CdS sample showed the highest rate constant of 0.0129 min$^{-1}$. The photodegradation data was well fitted by the pseudo-first order kinetic equation (Fig. 2b). When CdS was coupled with ZnO, the enhanced photocatalytic can be explained by the photosensitizer mechanism. Under the visible light irradiation, the photoexcited electrons in CB of CdS are transferred to the CB of visible-inert ZnO, and the photogenerated holes leave behind in the VB of the CdS as shown in Fig. 3. The charge transfer by 'photosensitizers' ensured separation of electron-holes and promoted the subsequent redox reactions leading to enhance the photocatalytic activity of the nanocomposite. If $O_v$-ZnO-CdS is a type II heterojunction, both photocatalysts can be excited under visible light irradiation. The excited electrons transfer to the CB of oxygen-vacancies enriched ZnO and further get trapped by $V_o$-level with lower energy [8][17]. The charge transfer by Type II heterojunction is not favorable for the generation of $·O_2^-$ and $·OH$ radicals, since the $V_o$-level potential is more positive compared to the standard redox potential for the formation of superoxide anion radical and VB potential is more negative compared to the redox potential for hydroxyl radical formation. Hence, by constructing a Z-scheme for the heterojunction, the generation and accumulation of electrons in the CB of g-$C_3N_4$ and holes in the VB of ZnO could be promoted, resulting in the generation of $·O_2^-$ and $·OH$ radicals. The p-benzoquinone (BQ), isopropanol (IPA) and EDTA-2Na were employed as the quenchers for $O_2^-$, $·OH$, and $h^+$, respectively [32]. The photocatalytic activity of $O_v$-ZnO-rGO-CdS nanocomposite was significantly decreased with the addition of IPA and EDTA-2Na scavengers (Fig. 2c), which confirms that $O_v$-ZnO-rGO-CdS nanocomposite is unfavorable for the formation of active species if the charge transfer conforms to the type II heterojunction. It further implied that the holes exhibited a preference for accumulation in the VB of ZnO, resulting in the subsequent generation of $·OH$ (Eq. 3-4). The proposed degradation mechanism of the photocatalyst can be summarized as follows [33]:

$$ZnO + h\nu \rightarrow h^+ + e^- \quad (1)$$

$$CdS + h\nu \rightarrow h^+ + e^- \quad (2)$$

$$e^-(CB\ of\ ZnO) \xrightarrow{rGO} combine\ with\ h^+ (VB\ of\ CdS) \quad (3)$$

$$h^+ (VB\ of\ ZnO) + OH^- \rightarrow ·OH \quad (4)$$

$$e^- (CB\ of\ CdS) + O_2 \rightarrow ·O_2^- \quad (5)$$

$$·O_2^- + ·OH + RhB \rightarrow Oxidation\ Products \quad (6)$$

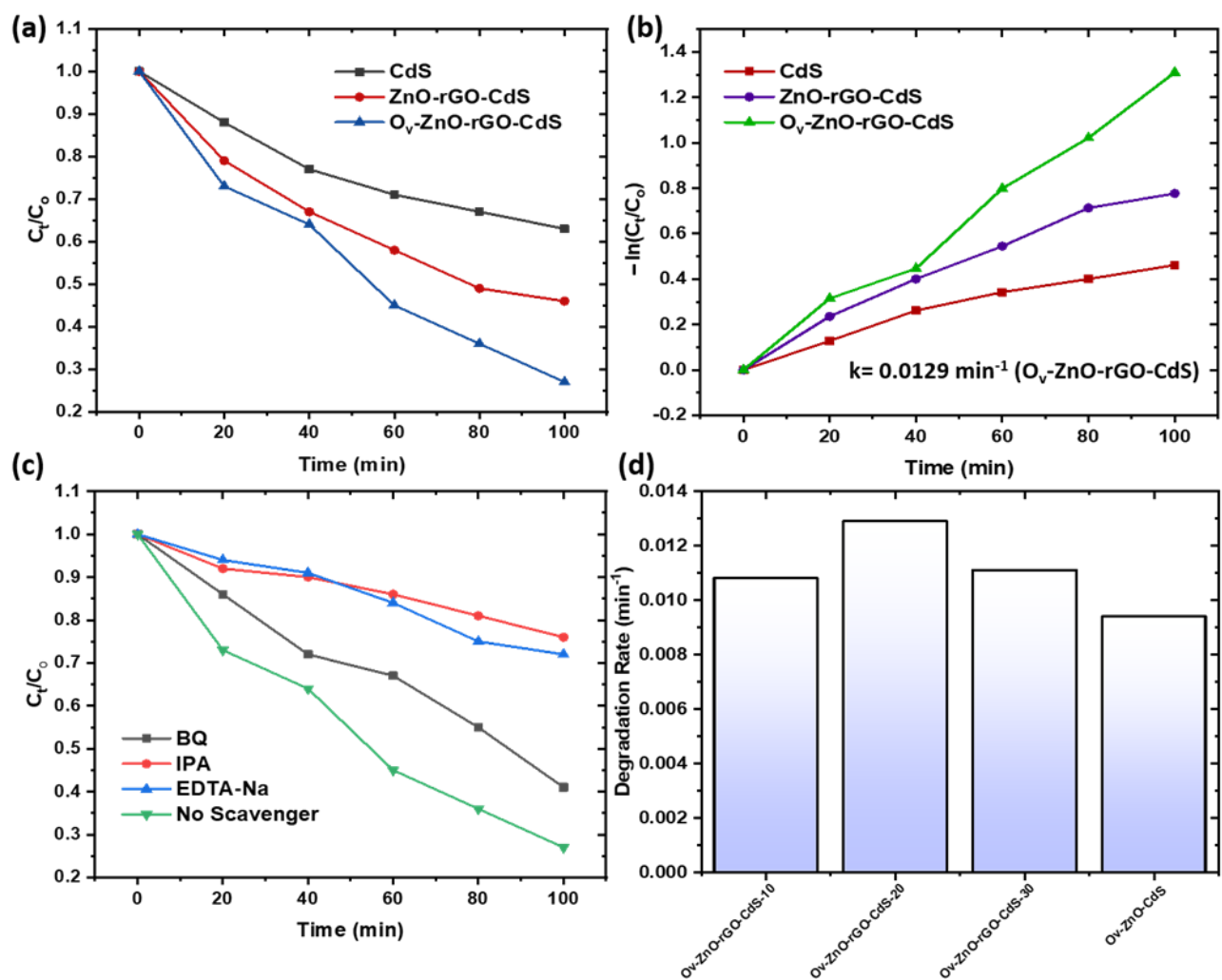

**Figure 2. (a)** Photocatalytic degradation of RhB where $C_t$ is the concentration after different light irradiation times and $C_o$ is the initial concentration. **(b)** Kinetic rate constants for the degradation of RhB **(c)** Reactive species trapping experiments of $O_v$-ZnO-rGO-CdS under visible light irradiation. **(d)** Degradation rate constants of $O_v$-ZnO-rGO-CdS nanocomposite at different wt.% CdS and in the absence of rGO.

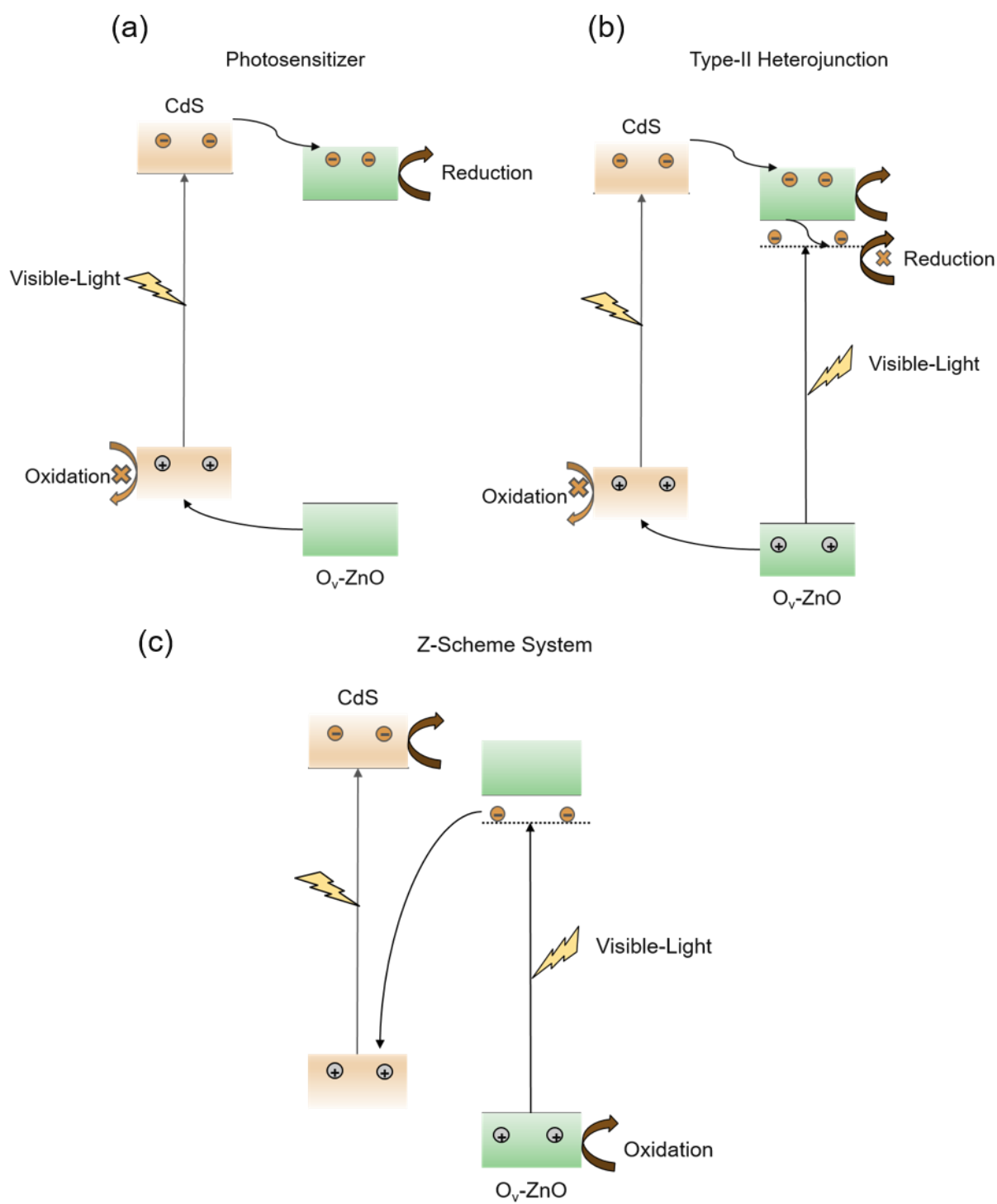

**Figure 3.** Schematic illustration of the charge-transfer in CdS coupled with $O_v$-ZnO under visible-light irradiation **(a)** Photosensitizer **(b)** Type-II heterojunction **(c)** Z-scheme.

rGO sheets with an appropriate Fermi level and high electronic conductivity act as an excellent channel for the recombination of photoexcited electrons from ZnO and holes from CdS by shortening the diffusion length for subsequent oxidation and reduction reactions [34]. The optimum rGO content was determined to 1:1, and a higher content leads to a decrease in the photocatalytic activity, probably due to a shielding effect [35]. $O_v$-ZnO-rGO-CdS heterostructures with varying amounts of CdS nanoparticles with 20 wt. % displayed the highest photocatalytic activity. As the amount of CdS increases, the reaction rate constant is increased since the CdS nanoparticles enhance the visible light absorbance and act as reduction sites. However, at 30 wt. % of CdS, activity decreased, as shown in Fig. 2d, which can be due to the overcoating of CdS on ZnO, that decreases the exposed ZnO surface as oxidative sites.

In ZnO-rGO-CdS heterostructure, the photocatalytic activity is contributed by the photosensitizer mechanism instead of the typical type-II heterojunction and Z-scheme. This leads to the weakening of the redox abilities of electrons and holes. In the ZnO-CdS heterostructure also, the migration of charge carriers is dominated by the photosensitizer mechanism. Furthermore, the apparent rate constant k of the $O_v$-ZnO-rGO-CdS photocatalyst is 1.37 times greater than that of $O_v$-ZnO-CdS. This result indicates that the presence of rGO could effectively improve the charge transfer process between ZnO and CdS due to its high electron mobility and enhance the stability of CdS against photocorrosion. In $O_v$-ZnO-CdS heterostructure, transfer by type II is not favorable, like $O_v$-ZnO-rGO-CdS, and hence the transfer of charge carries occurs by Z-scheme. However, interfacial recombination of photoinduced electrons trapped in the Vo level of ZnO with photoinduced holes in the VB of CdS is not that highly efficient due to the absence of the redox mediator in the form of rGO, and this explains the reason behind the reduction in the photocatalytic activity for Rh degradation.

## 4. Conclusion

This study focuses on enhancing the photocatalytic properties of ZnO by implementing an oxygen defects-mediated Z-scheme mechanism. The successful synthesis of a defect-mediated ZnO-CdS nanocomposite with rGO as an electron mediator was demonstrated, which effectively addresses the limitations observed in traditional ZnO photocatalysts. By leveraging the synergistic effects of $O_v$-ZnO, CdS, and rGO, we achieved enhanced photostability and visible-light-induced photocatalysis properties. This work sheds light on the defect-dependent interfacial mechanism in heterostructure nanocomposites, offering valuable insights for the development of high-performance ZnO-based photocatalysts. The integration of the defect-mediated Z-scheme mechanism and rGO as a solid-state electron mediator presents a promising approach to overcome the limitations observed in traditional ZnO photocatalysts. The proposed mechanism provides valuable insights into the role of defects in interfacial charge transfer within heterostructure nanocomposites. The optimal rGO content in $O_v$-ZnO-rGO-CdS was found to be 1:1, with higher contents decreasing photocatalytic activity possibly due to a shielding effect. The highest photocatalytic activity was achieved with 20 wt. % CdS, enhancing visible light absorption and acting as reduction sites, resulting in an increased reaction rate constant. However, at 30 wt. % CdS, activity decreased due to overcoating, reducing the exposed ZnO surface. rGO improved charge transfer between ZnO and CdS, enhancing CdS stability against photocorrosion. The apparent rate constant (k) of $O_v$-ZnO-rGO-CdS was 1.37 times higher than $O_v$-ZnO-CdS, indicating the beneficial effect of rGO in facilitating efficient charge transfer. Based on the investigation, the role of the oxygen vacancies in ZnO as a mediator of the Z-scheme separation of the heterojunction could be summarized by the following two aspects. On the one hand, the $O_v$ generated isolated level below the CB of ZnO caused a bandgap reduction to make the heterojunction more visible-light absorbable. On the other hand, under visible light, the photo-induced electrons trapped in the $O_v$ could be directly recombined with the photo-induced holes in the VB of CdS through rGO. This work provides a new paradigm for the design of an efficient visible-light-activated Z-scheme system for energy and environmental applications.